\documentstyle[aps,epsbox,prl,twocolumn]{revtex}
\begin{document}

\title{Dynamical process for switching between the metastable ordered magnetic state and the nonmagnetic ground state in photoinduced phase transition
}

\author{Masamichi Nishino and Seiji Miyashita}

\address{Department of Applied Physics, Graduate School of Engineering,\\
The University of Tokyo, Bunkyo-ku, Tokyo, Japan}

%\wideabs{
\maketitle
\begin{abstract}                % DON'T CHANGE THIS LINE
We propose a dynamical mechanism of the two-way switching between
the metastable state and the stable state, which has been found in experiments of photoinduced reversible magnetization and photoinduced structural phase
transition.
We find that the two-way switching with a non-symmetry breaking
perturbation such as illumination is possible only in systems with appropriate parameters. We make it clear that the existence of two time scales in the dynamical process is important for the two-way switching.

\end{abstract}

%}

%\begin{multicols}{2}
%\narrowtext

%\section{Introduction}
%\markboth{EFFECT OF SPIN PUMPING BY ILLUMINATION}{Introduction}

Dynamics of the switching between the metastable state and
the stable state is one of the most interesting topics of phase transitions.
Photoinduced phase transition (PIPT) has currently attracted interests in
the condensed matter
physics~\cite{Koshihara1,Sci96,J.Elec97,Miyano,Koshihara2,Ogawa}.
In spin-crossover complex several such phenomena have been
reported~\cite{Koshihara2,Hauser,Gutlich}. 
PIPT is accompanied by a structural phase transition as observed in TTF-CA~\cite{Koshihara2} or by a magnetic phase transition as observed in Co-Fe complex~\cite{Sci96,J.Elec97}.
Because the energy level of the low spin state of such a complex gets cross
to that of the high spin state, the spin state can be changed between these
states by stimulus due to changes of temperature, pressure, photoexcitation, etc.
Among them, the cobalt-iron prussian blue analogs~\cite{Sci96,J.Elec97} have
been reported to be able to switch from the paramagnetic state to the
ferrimagnetic state (or vice versa) by visible (or near-IR) illumination.

In order to study the switching mechanism, we consider a system which has a
long-lived metastable ferromagnetic excited state.
The transition process from the metastable magnetic state to the stable paramagnetic state can be realized by the nucleation
process because the bulk free energy of the stable state is lower than that
of the initial metastable state~\cite{Rikvolt}.
Hereafter we refer to the metastable state as MS and to the stable state as SS.
On the other hand, switching from SS to MS conflicts with the thermodynamic
stability and it is considered to be difficult to realize as far as we apply only symmetric
disturbance, e.g., illumination, although it is easy by using a
uniform field which causes the symmetry breaking.
Our interests is how the two-way switching is realized in a symmetric applied field.
In our previous study~\cite{Nishino}, photon's effect is attributed to the
change of renormalized parameters $(D,T)$, where we chose these parameters
in the phase diagram from a viewpoint of phenomenology.

In this letter we consider the condition for the two-way switching in terms of the transition probabilities among microscopic states.
We find that the two-way switching cannot be realized when we introduce only a
spin-pumping effect by photon.
In addition to the pumping effect of photon,
a suitable relation between transition probabilities among elementary
processes is
necessary to realize the two-way switching. Here we will extend the 
Glauber dynamics~\cite{Glauber} and introduce a more complicated
relaxation process
where the switching (SS $\rightarrow$ MS) can be realized. Using this model
we will study characteristics of the system which can show the two-way
photoinduced
transition.

As a typical model which shows the first-order phase transition, we adopt the
Blume-Capel (BC) model~\cite{BC1,BC2} in the simple cubic lattice,
\begin{equation}
{\cal H}= -J \sum_{\langle i, j \rangle} S_iS_j + D \sum_{i}{S_i}^2,
\end{equation}
 where $S_i$=$\pm 1$ or 0  and $\langle i, j \rangle$ denotes
the nearest-neighbor pairs.
Here $S_i=\pm 1$ represents the high spin state with the magnetization $\pm
1$,
respectively, and $S_i=0$ represents the low spin state which is nonmagnetic.
The excitation energy from the low spin state to the high spin state is
$D(>0)$. We take $J(>0)$ as a unit of energy and set $k_{\rm  B}=1$.

\begin{figure}
\begin{center}
\epsfile{file=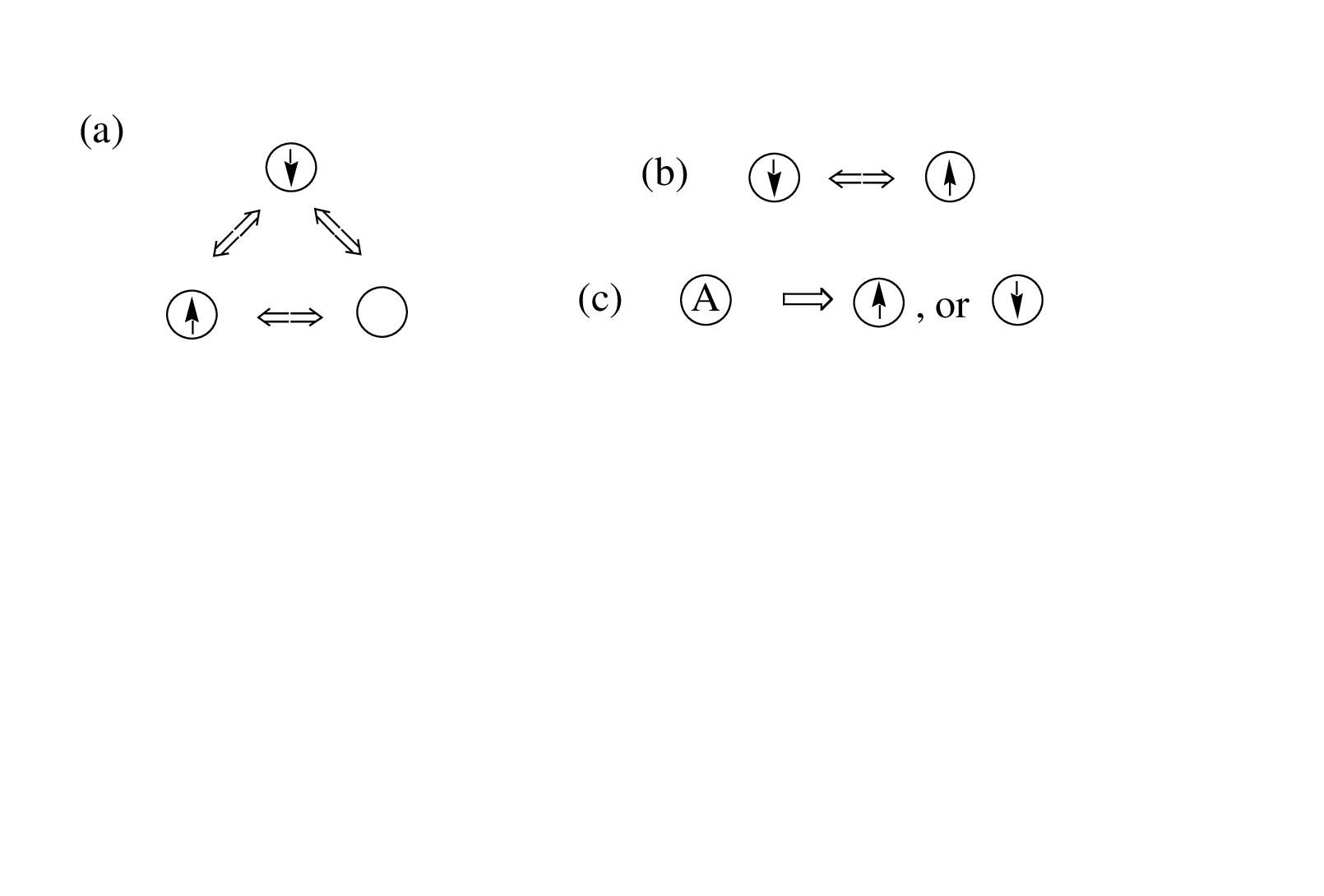,scale=0.3}
\end{center}
\vspace{-2.5cm}
\caption{Transitions among states. The circle is the $k$th site.
$\uparrow$ and $\downarrow$ in the circle denote up spin state and down spin
state,
respectively. Blank circle denotes a nonmagnetic state.
(a) transition process among all the possible local states.
(b) transition process between the high spin states $S=\pm1$.
(c) transition process by pumping. A in the circle denotes
any state ($S_k$=1, $-$1 or 0).}
\label{Fig.1_operator}
\end{figure}

Here we adopt a stochastic dynamics with multi-time scales as follows.
Let $P(S_1 \cdots S_k \cdots S_N, t)$ be the
probability of the state $\{S_1 \cdots S_k \cdots S_N \}$ at time $t$.
Using the transition probability per a unit time $W_k(S_k \rightarrow S_k')$
at the $k$-th site  (Fig. \ref{Fig.1_operator} (a)), the master equation is
generally given
by
\begin{eqnarray}
&&\frac{d}{dt} P(S_1 \cdots S_k \cdots S_N, t) =   \label{eq:glauber} \\
&&-\sum_k \sum_{S_k \neq S_k'} W_k(S_k \rightarrow S_k')  P(S_1 \cdots S_k
\cdots S_N, t) \nonumber \\
&&+\sum_k \sum_{S_k \neq S_k'} W_k(S_k' \rightarrow S_k) P(S_1 \cdots S_k'
\cdots S_N, t).  \nonumber
\end{eqnarray}
Here, we assume that the time scale
of the transition between $S=\pm1$ is much faster than that between $S=0$
and $|S|=1$, because the former is a simple spin flip process while the
latter is accompanied by a structural change.
Representing this situation, we introduce two transition processes:
The first one is the standard Glauber type evolution,
\begin{eqnarray}
W_k(S_k \rightarrow 1) &=&\frac{\exp(y)}{2\cosh y + \exp(\beta D)} 
\nonumber \\
W_k(S_k \rightarrow -1) &=&\frac{\exp(-y)}{2\cosh y + \exp(\beta D)}  \label{eq:detail_glauber} \\
W_k(S_k \rightarrow 0) &=& \frac{\exp(\beta D)}{2\cosh y + \exp(\beta
D)}  \nonumber
\end{eqnarray}
where $S_k=\pm1, \; 0$(Fig. \ref{Fig.1_operator} (a)) and $y=\beta
J\sum_{\rm{NN}}S_i$. Here $\sum_{\rm{NN}}$ denotes the summation over the
nearest neighbor sites of $k$. We express this process by $\frac{d}{dt}
\mbox{\boldmath$P$}(t)=\hat{L}_{\rm G}(t) \mbox{\boldmath$P$}(t)$.
The second one is a process of spin flip, $\hat{L}_{\rm S}$, (Fig. \ref{Fig.1_operator}
(b)),
\begin{eqnarray}
W_k(\pm1\rightarrow 1) &=& \frac{\exp(y)}{2\cosh y} 
\nonumber \\
W_k(\pm1\rightarrow -1) &=& \frac{\exp(-y)}{2\cosh y} 
 \label{eq:trans} \\
W_k(S_k \rightarrow S_k') &=&0  \; \; \; {\rm (the \; other \; weights)}
\nonumber
\end{eqnarray}
We study a combination of these two processes,
\begin{equation}
\hat{L}(t)=p_{\rm G} \hat{L}_{\rm G}(t)+ p_{\rm S} \hat{L}_{\rm S}(t).
\end{equation}
Here the parameters $p_{\rm G}$ and $p_{\rm S}$ ($p_{\rm G}+p_{\rm S}=1$) describe the relative
time scale of the processes.
When $p_{\rm G}/p_{\rm S}$ is small, the relaxation among the high spin 
states ($S=\pm1$) occurs more quickly than that between high and low spin 
states ($|S|=1$ and 0).
Because both $\hat{L}_{\rm G}$ and $\hat{L}_{\rm S}$ satisfy the detailed balance,
it is warranted that with any combination of $p_{\rm G}(\neq 0)$ and $p_{\rm S}$ any
initial state goes to the equilibrium state.
Here we think that the ratio of $p_{\rm G}$ and $p_{\rm S}$ is given as an inherent
property of individual materials.
Under the process $\hat{L}_{\rm S}$, the pumped spins tend to align before the
system relaxes back to the state of $S=0$. When the adequate balance of these
two processes is realized, we expect that the reversible switching of the
magnetization by illumination is possible.

Next we consider a process $\hat{L}_{\rm P}$ for illumination.
During the illumination, the pumping process is included.
Photon excites a site which is either nonmagnetic or magnetic to a
magnetic site.
It is a natural assumption that the transition to up spin or to down spin
has an equal probability and we take the transition probabilities (Fig.
\ref{Fig.1_operator} (c)),
\begin{eqnarray}
W_k(S_k \rightarrow \pm1) &=& \frac{1}{2} .
\end{eqnarray}
 The time evolution during the illumination is given by
\begin{equation}
\hat{L}(t)=p_{\rm G} \hat{L}_{\rm G}(t) +p_{\rm S} \hat{L}_{\rm S}(t)+p_{\rm P} \hat{L}_{\rm P}(t),
\end{equation}
where $p_{\rm G}+p_{\rm S}+p_{\rm P}=1$. The magnitude of $p_{\rm P}$ represents the strength of the
excitation. Indeed in the experiments~\cite{Sci96,J.Elec97}, two kinds of
illuminations with different frequencies are used to realize the switching of magnetization.

We look for a suitable set of  $p_{\rm G}$, $p_{\rm S}$ and $p_{\rm P}$ with which the system can be switched between MS and SS. Let us study the dynamics of the system first
within a mean field theory (MFT). We will then check the stability of the 
results of MFT by a Monte Carlo (MC) method.

\begin{figure}
\begin{center}
\epsfile{file=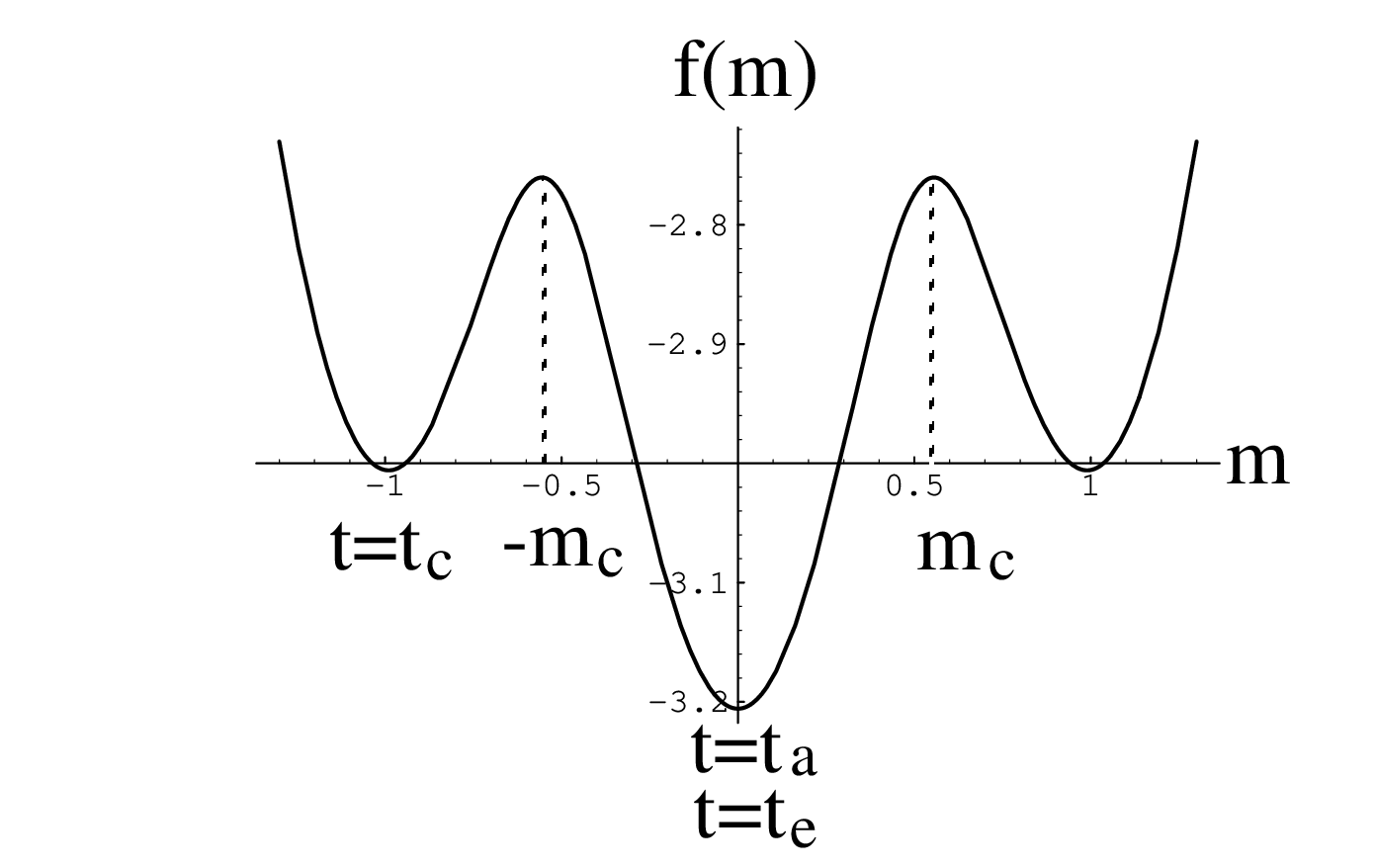,scale=0.4}
\end{center}
\caption{
Mean field free energy at the equilibrium state for $(D,T)=(3.2,0.6)$.
}
\label{Fig.2_free_energy_eq}
\end{figure}

In MFT, Eq. (\ref{eq:glauber}) is reduced to an equation
of $p_1(t)$, $p_{-1}(t)$, and $p_0(t)$, which are respectively the
probabilities for up spin state, down spin state, and nonmagnetic state on a
site at time $t$.
Furthermore, the equations of $p_1(t)$ and $p_{-1}(t)$ are reduced to an
equation of the magnetization, and the equation is found to be expressed in
the form of the Van Hove's phenomenological equation.
\begin{equation}
\frac{d}{dt} m = \dot{p}_1(t)-\dot{p}_{-1}(t)=-b \frac{d f(m,t)}{dm}.
\label{eq:m_t}
\end{equation}
This $f(m,t)$ is regarded as the effective free energy, which is useful
to  visualize the thermodynamic force at a given time $t$.
We obtain $f(m,t)$ by integrating the right hand side of Eq. (\ref{eq:m_t})
as to $m$.
\begin{eqnarray}
\lefteqn{f(m, t)  =  {1\over2} Jzm^2  - p_{\rm G} {1\over{\beta}} \; {\ln[2\cosh
({\beta}Jzm)
+\exp({\beta}D)] } } \nonumber \\
 &-& p_{\rm S} {1\over{\beta}}  \{ \; \ln[\cosh ({\beta}Jzm)]
(1-p_0(t=0)\exp(-(p_{\rm G}+p_{\rm P}) t) ) \nonumber \\
 &-& \frac{p_{\rm G}}{p_{\rm G}+p_{\rm P}} (1-\exp(p_{\rm G}+p_{\rm P}) t) ( \ln[\cosh ({\beta}Jzm)]
\label{eq:td_free_energy}  \\
&-& \ln[2 \cosh ({\beta}Jzm) +\exp(\beta D) ] ) \}. \nonumber
\end{eqnarray}
In the case of no illumination
($p_{\rm P}=0$), the above free energy gives the equilibrium free energy
$f_{\rm eq}(m)$ in the limit $t \rightarrow \infty$ regardless of the parameters $p_{\rm G}(\neq 0)$ and $p_{\rm S}$.
In Fig. \ref{Fig.2_free_energy_eq} $f_{\rm eq}(m)$ for $(D,T)=(3.2,0.6)$ is shown.

Because of the existence of metastable valleys, some initial states may
relax to the MS instead of SS.
In the case of $p_{\rm S}=0$ ($p_{\rm G}=1$), $f(m,t)$ is independent of time and takes the form of $f_{\rm eq}(m)$ in Eq.(\ref{eq:td_free_energy}).
When the value of magnetization is larger than the critical value $m_c(=\pm 0.5557)$, the state relaxes to MS.
Thus, in order to make the system to relax to MS after the illumination, 
the magnetization in the illumination must increase beyond $m_c$.
It is found however that the magnetization of the stationary state in
the illumination is much smaller than $m_c$ for any value of $p_{\rm P}$. Therefore it is impossible to cause the transition SS $\rightarrow$ MS when $p_{\rm S}=0$.

\begin{figure}
\begin{center}
\epsfile{file=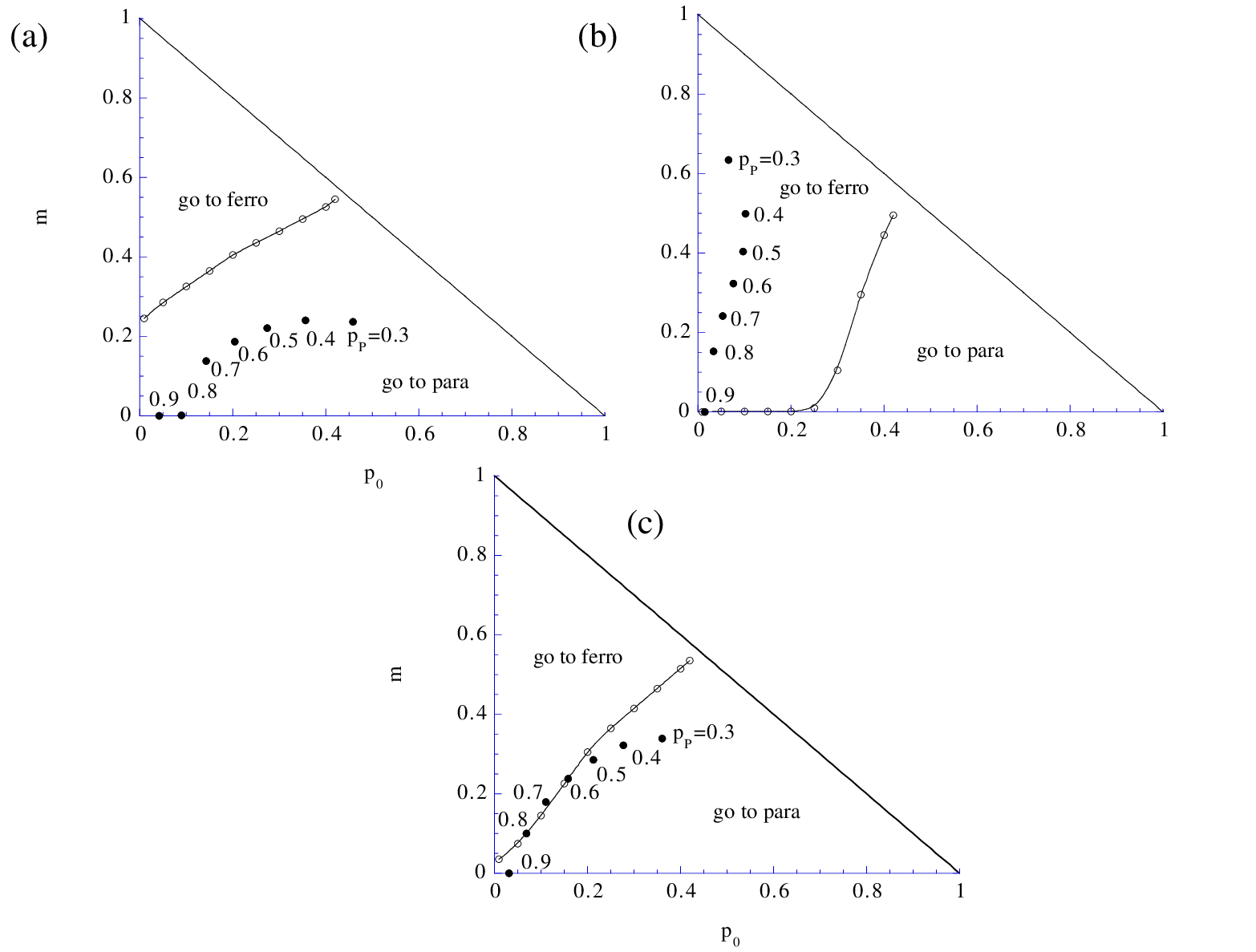,scale=0.4}
\end{center}
\caption{Critical line linking the open circles in the relaxation process ($p_{\rm P}=0$) and
the values ($p_0$, $m$) of the stationary state in the illumination are also plotted by the closed
circle for several values of $p_P$.
(a)$p_{\rm G}/p_{\rm S}=0.667$
(b)$p_{\rm G}/p_{\rm S}=0.163$
(c)$p_{\rm G}/p_{\rm S}=0.429$
}
\label{Fig3_critical_line}
\end{figure}

Next we consider the case of $p_{\rm S}\neq0$, where the process of spin ordering
is expected to be enhanced. 
In spite of the case of $p_{\rm S}=0$, the form of $f(m,t)$ depends on $t$ and $p_0(t=0)$.
In Fig. \ref{Fig3_critical_line}, we show a kind of phase diagram of the
initial values ($p_0$, $m$) for the cases of
(a)$p_{\rm G}/p_{\rm S}=0.667$, (b)$p_{\rm G}/p_{\rm S}=0.163$ and (c) $p_{\rm G}/p_{\rm S}=0.429$.
Because of the condition $p_1+p_{-1}+p_0=1$ and $0\leq p_1, p_{-1}, p_0 \leq
1$, $p_0+m \leq 1$ must be satisfied.
The open circles denote the boundary between
the region of ($p_0$, $m$) from where the state goes to MS and the region from where the state goes to SS.
There the values ($p_0$, $m$) of the stationary state of the illumination
process ($p_{\rm P}\neq 0$) are also plotted by closed circles for various values of $p_{\rm P}$ keeping the ratio of $p_{\rm G}/p_{\rm S}$. That is, in the case of $p_{\rm P}=0.5$ and $p_{\rm G}/p_{\rm S}=0.667$,
$p_{\rm G}=0.2$ and $p_{\rm S}=0.3$.

If a closed circle for a value of $p_{\rm P}$ is located below the critical line linking open circles, the initial state with the values ($p_0$, $m$) denoted by this closed circle relaxes to SS (para).
By the illumination the system becomes to the state denoted by this closed circle, and then after the illumination it relaxes to SS. 
In the case of (a) the point ($p_0$, $m$) of the stationary state with any value of $p_{\rm P}$ is 
located below the critical line as shown in Fig.3 (a), and thus all 
states after the illumination give the initial state for SS.
On the other hand, if a closed circle for a value $p_{\rm P}$ is located above the critical line, the state relaxes to MS.  
By this illumination the system becomes to the state denoted by this closed circle, and then after the illumination it relaxes to MS (ferro). 
In the case of (b) the point of the stationary state with any value of $p_{\rm P}$ is located above the critical line as shown in Fig.3 (b),  and thus all states after the illumination give the initial state for MS.
In order to realize the two-way switching  (MS$\rightleftharpoons$SS)
the system has to have both cases for different values of $p_{\rm P}$.
That is, the closed circles are located on both sides of the critical line as shown in Fig. 3(c).

With the present observation we find that the two-way switching is not
possible in every system but in only limited materials which have
a peculiar property that the boundary (open circles)
crosses with the line of closed circles.  On the other hand,
we also find that such situation can be actually exist in some appropriate
condition.

\begin{figure}
\begin{center}
\epsfile{file=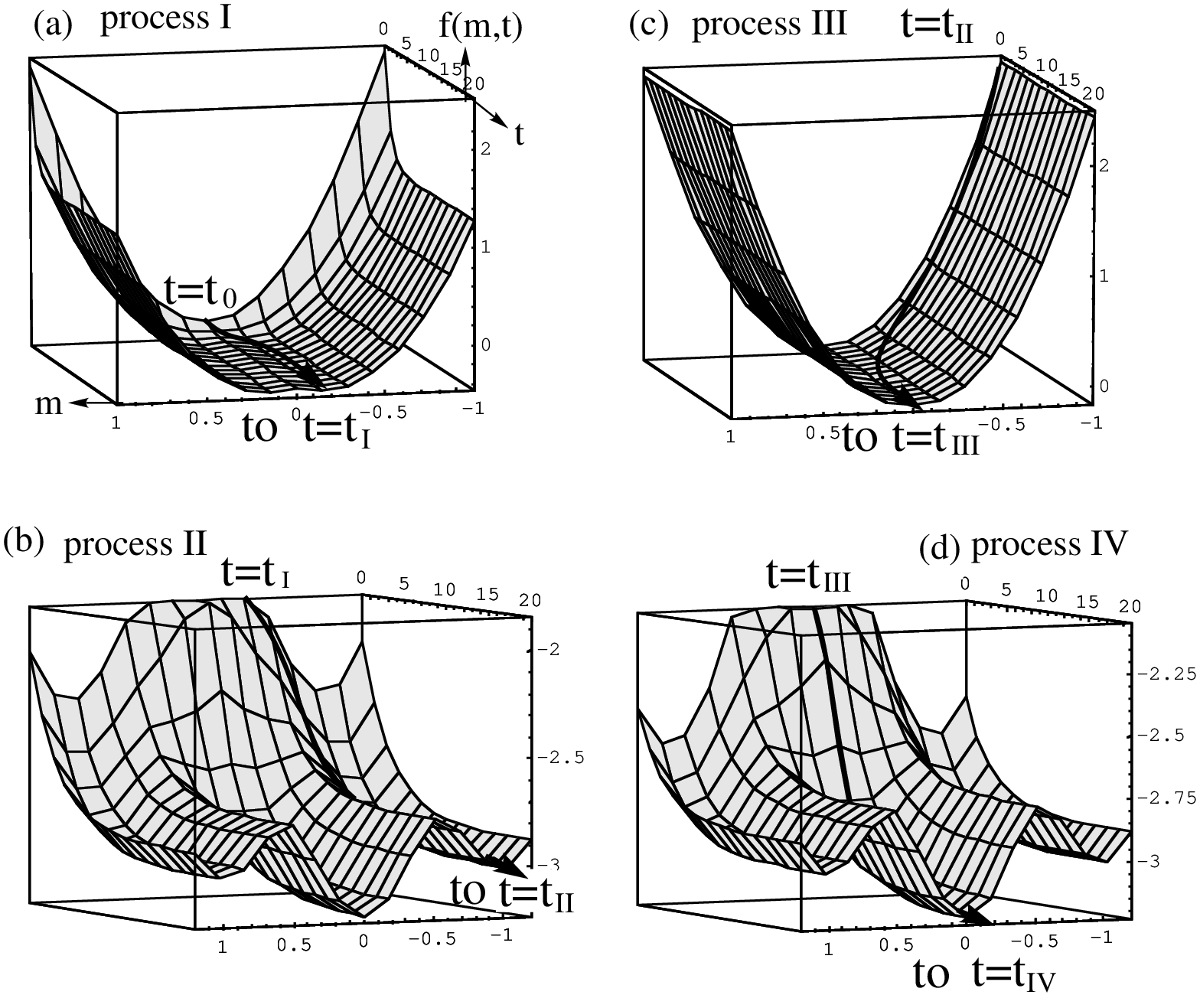,scale=0.42}
\end{center}
\caption{
Time evolution of $f(m,t)$ fixing the ratio $p_{\rm G}/p_{\rm S}=0.429$.
(a) process I, $t_0 <t<t_{\rm I}$ ($p_{\rm P}=0.7$), (b) process II, $t_{\rm I}
<t<t_{\rm II}$, (c) process III, $t_{\rm II} <t<t_{\rm III}$ ($p_{\rm P}=0.9$),
and (d) process IV, $t_{\rm III} <t<t_{\rm IV}$. The thick arrows show the
change of the magnetization.
The change of $f(m,t)$ is shown for $t_{\rm a}<t<t_{\rm a}+20$ (a=0, I, II
and III) which corresponds to the first 1/5 of the process because $f(m,t)$
changes little after $t_{\rm a}+20$. Here the sign of magnetization is not
essential.
}
\label{Fig4_free_energy}
\end{figure}

Now we demonstrate how the switching of the magnetization is realized.
According to the results in Fig. 3,  we choose the ratio $p_{\rm G}/p_{\rm S}$=0.429.
As an initial state we take the equilibrium state of the parameter $(D,T)=(3.2,0.6)$ where $m=0$ and $p_0=0.99$. 
We make the system in the illumination of $p_{\rm P} = 0.7$ for a period $t_0 <t< t_{\rm I}$ (process I),
and then we turn off the illumination and let the system relax, i.e.
$p_{\rm P} = 0$ for a period $t_{\rm I} <t<t_{\rm II}$ (process II). 
The change of the effective free energy $f(m,t)$ in the process I is shown
in Fig. \ref{Fig4_free_energy} (a).
At the end of the process $f(m,t)$ almost reaches the form of its stationary state
where $f(m,t)$ has two minima and $m=\pm 0.18$.
In the second process $f(m,t)$ changes as shown in Fig. \ref{Fig4_free_energy} (b).
Because the final state of the process I $(p_0,m)=(0.11,0.18)$ locates in the region to go to MS (ferro), the system relaxes to the metastable state. 
Thus we have the switching SS $\rightarrow$ MS.
We see that the magnetization increases with time in the process II.

%\end{multicols}
%\widetext

\begin{figure}
\begin{center}
\epsfile{file=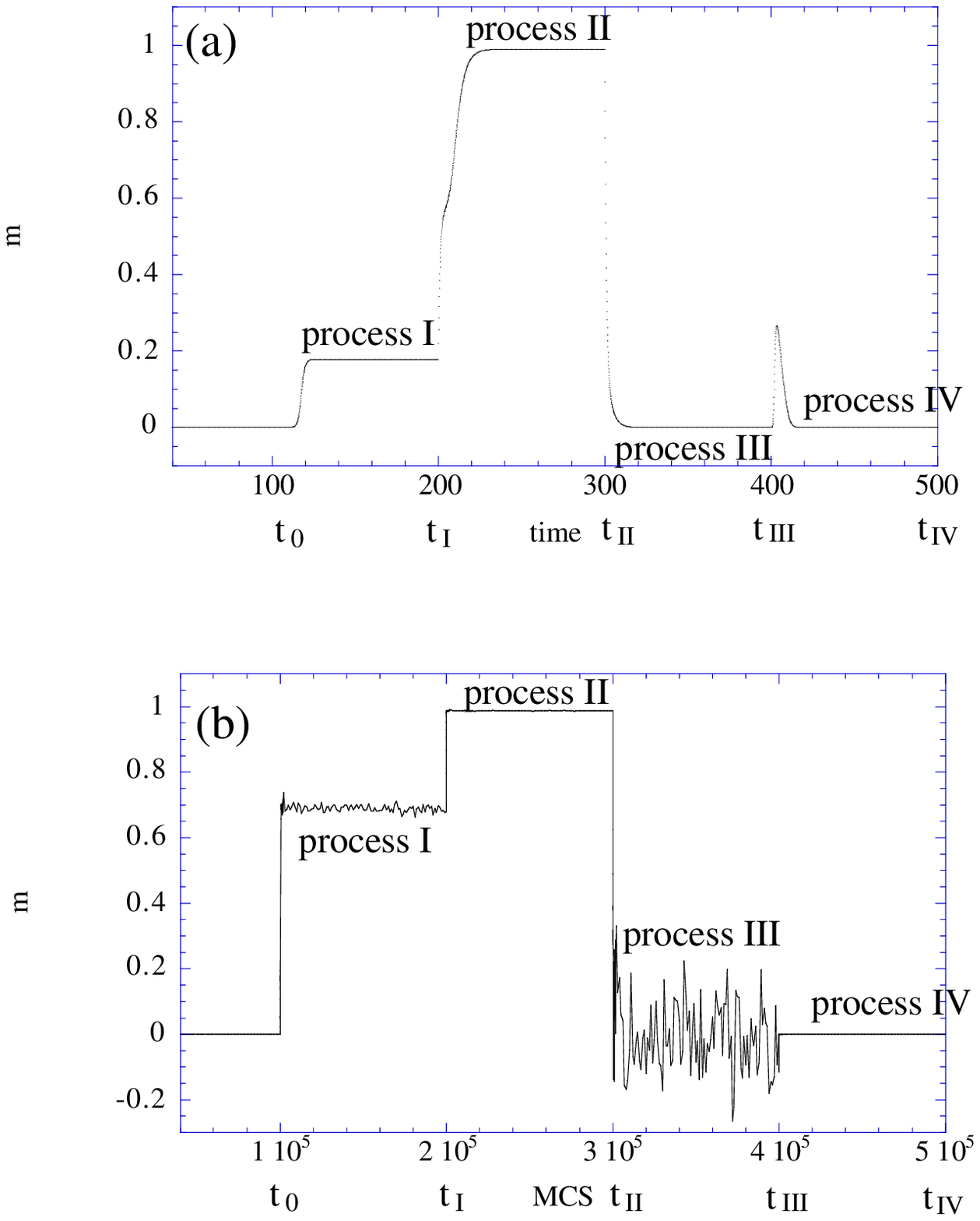,scale=0.45}
\end{center}
\caption{(a)Time evolution of magnetization per site for the parameter set
$p_{\rm G}/p_{\rm S}=0.429$. $p_{\rm P}=0.7$ and $p_{\rm P}=0.9$ are adopted as the first and the
second illumination respectively in the mean field theory.
(b) Time evolution of magnetization per site for the parameter set
$p_{\rm G}/p_{\rm S}=0.289$. $p_{\rm P}=0.096$ and $p_{\rm P}=0.24$ are adopted as the first and the
second illumination respectively in the Monte Carlo method.}
\label{Fig5_mag}
\end{figure}

Next, let us consider a procedure for the switching MS $\rightarrow$ SS.
Now we illuminate the system with another light of
$p_{\rm P}=0.9$ for a period $t_{\rm II} <t<t_{\rm III}$ (process III), 
where $f(m,t)$ changes as shown in Fig. \ref{Fig4_free_energy} (c).
The $f(m,t)$ at the final state ($t=t_{\rm III}$) is parabolic and the system has almost no magnetization.
Thus after turning off the light the system relaxes to the stable
paramagnetic state as shown in Fig. \ref{Fig4_free_energy} (d) (process IV).
The time evolution of magnetization of the whole processes is shown  in Fig.
\ref{Fig5_mag} (a). In the process III  we could choose $p_{\rm P}$ to be 0.4 instead of 0.9 to realize MS$\rightarrow$SS because the closed circle 
for $p_{\rm P}=0.4$ is also located below the critical line.

We have shown that the two-way switching is realized by the processes 
I $\sim$ IV in MFT. 
In order to check whether the switching process is stable against fluctuation,
we study the processes by a Monte Carlo (MC) method.
Of course quantitatively MC and MFT gives different results.
But, qualitatively the same features of the dynamics as obtained in MFT are also found in MC. Namely, when $p_{\rm S}=0$, the switching SS $\rightarrow$ MS is not observed for any value of $p_{\rm P}$.
On the other hand we can realize the two-way switching by tuning the parameters
as $p_{\rm G}/p_{\rm S}=0.289$, $p_{\rm P}=0.096$ for the process I and  $p_{\rm P}=0.24$ for the process III.
The time evolution of magnetization per site is shown in Fig. \ref{Fig5_mag}
(b) ( simple cubic lattice with 1000 sites in periodic boundary condition).
Strictly speaking, in MC the true stationary state of the process II is the
paramagnetic state, but the relaxation MS $\rightarrow$ SS is
very slow and thus practically the same behavior as in MFT is reproduced.  
We confirmed that this switching of magnetization can be repeated reliably,
and we conclude that the two-way switching provided by the processes I - IV
is stable even in systems with short range fluctuations. 
In the MC simulation, the switching dynamics depends very sensitively on the 
ratio of parameters  as well as in the MFT case.

In the evolution ${\hat L_{\rm P}}$,
we considered that illumination causes the low spin state to be excited to the high 
spin state. However, illumination may also cause the high spin state to change to
the low spin state, which disturbs the development of magnetization and
gives a negative effect on the switching SS $\rightarrow$ MS. 
As the extreme case we consider that illumination brings  one of three
states with the equal probability regardless of the initial state, i.e.,
$W_k(S_k \rightarrow S_k')=\frac{1}{3}$ for any $S_k$ and $S_k'$.  
Although it is difficult  to grow a magnetic droplet in this case,
we found that the switching (SS $\rightleftharpoons$ MS) is still possible.

We studied dynamical processes for photoinduced two-way switching between
the ordered state and the disordered state in a master equation approach.
We have found that a peculiar condition is necessary between the transition
probabilities. That is, only when the transition probability between the high and low spin states
and that between the high spin states satisfy a special ratio, 
the two-way switching by using different illuminations  is possible.
This implies that only materials which satisfy such conditions can show the 
photoinduced switching phenomena.

The present work was supported by Grant-in-Aid for Scientific
Research from Ministry of Education, Science, Sports and
Culture of Japan.
M. N. was also supported by the Research Fellowships of the Japan
Society for the Promotion of Science for Young Scientists.

%\end{multicols}

\end{document}